\documentstyle[12pt]{article}
\newcommand{\be}{\begin{equation}}
\newcommand{\ee}{\end{equation}}
\newcommand{\bea}{\begin{eqnarray}}
\newcommand{\eea}{\end{eqnarray}}
\newcommand{\bc}{\begin{center}}
\newcommand{\ec}{\end{center}}
\newcommand{\vac}{\mid 0\rangle}
\newcommand{\x}{{\vec{\rm x}}}
\newcommand{\ka}{{\vec{\rm k}}}
\newcommand{\y}{{\vec{\rm y}}}
\newcommand{\q}{{\vec{\rm q}}}
\newcommand{\p}{{\vec{\rm p}}}

\newcommand{\rv}{{\vec{\rm r}}}

\newcommand{\s}{{\vec{\rm s}}}

\newcommand{\jv}{\vec{J}}
\newcommand{\kap}{\vec{\kappa}}
\newcommand{\di}{\displaystyle \int }

\newcommand{\JJ}{{\cal J}}
\newcommand{\PP}{{\cal P}}
\newcommand{\Pop}{{\mbox{\bf P}} }

\newcommand{\G}{{\mbox{\bf G}} }
\newcommand{\A}{\tilde{A}}
\newcommand{\B}{\tilde{B}}
\newcommand{\C}{\tilde{C}}
\newcommand%
{\E}%
[1]%
{E_2({\cal P},#1)}%
\newcommand%
{\I}%
[1]%
{{I^{{\cal P}\pm}_{#1}(q)}}%
\newcommand%
{\VF}%
[1]%
{\Phi^{\pm}_{{\cal P}q}(#1)}%
\newcommand%
{\vf}%
[2]%
{\phi^{\pm (#1)}_{{\cal P}q}(#2)_{\alpha\beta}}%
\newcommand%
{\J}%
[1]%
{{\rm J}^{\cal P}_{#1}(\sigma)}%
\newcommand%
{\CC}%
[1]%
{{{\cal C}^{{\cal P}\pm}_{#1}(q)}}%
\newcommand%
{\DD}%
[1]%
{{{\cal D}^{{\cal P}}_{#1}(\mp iq)}}%
\newcommand%
{\TT}%
[2]%
{{{\cal T}^{(#1)\pm}_{{\cal P}}(q;#2)}}%
\newcommand%
{\D}%
[1]%
{D^{\cal P}_{#1}(\sigma)}%

\begin{document}
\begin{center}
{\Large \bf Fine-Tuning Renormalization and Two-particle States in
Nonrelativistic Four-fermion Model}
\end{center}
\begin{center}
A.N.Vall, S.E.Korenblit, V.M.Leviant, A.V.Sinitskaya, A.B.Tanaev.
\footnote{This work is partially supported by RFFR N 94-02-05204.} \\
Irkutsk State University, 664003, Gagarin blrd, 20, Irkutsk, Russia.
\footnote{E-mail KORENB@math.isu.runnet.ru }
\end{center}
\begin{abstract}
{\small
Various exact solutions of two-particle eigenvalue problems for
nonrelativistic contact four-fermion current-current interaction are
obtained.
Specifics of Goldstone mode is investigated. The connection between a
renormalization procedure and construction of self-adjoint extensions
is revealed.}
\end{abstract}
\begin{center}
{\bf 1. Introduction }
\end{center}

Models with contact four-fermion interaction are considered in a
wide range of problems both in solid medium and in quantum field theory.
It is well known that in quantum field theory such interaction is
nonrenormalizble in the frames of conventional perturbation approach.
In the present work we give explicit solutions for two-particle eigenvalue
problem clarifying the meaning of renormalization in these
models. We show the different kinds of the solutions correspond to
different renormalization schemes which imply different self-adjoint
extensions for Hamiltonian in two-particle sector.

\begin{center}
{\bf 2. Contact four-fermion models }
\end{center}

Let us consider the following Hamiltonians:
\bea
&& H_1 =\int d^3x\left[\Psi^{\dagger}_{\alpha}(x)\,E({\Pop})\,\Psi_{\alpha}
(x) -\frac{\lambda}{8}\ \chi^{\dagger}(x)\ \chi(x)\right];
\quad\mbox{ (Toy Model)}
\label{H1} \\
&& H =\int d^3x\left\{\Psi^{\dagger a}_{\alpha }(x)\,{\cal E}({\Pop})\,
\Psi^a_{\alpha }(x) -\frac{\lambda}{4}\left[S^2(x)-
{\jv}^2(x)\right]\right\}.
\label{H2}
\eea
defining: $ x=(\x,t=x_0)$, $\;{\Pop}_{\rm x}=-i\vec{\nabla}_{\rm x}$,
$\;\epsilon_{\alpha\beta}=-\epsilon_{\beta\alpha}$,
$\;\chi(x)=\epsilon_{\alpha\beta}\,\Psi_{\alpha }(x)\,\Psi_{\beta}(x)$,
\bea
&& S(x)=\Psi^{\dagger a}_{\alpha }(x)\,\Psi^a_{\alpha }(x);\;\;\;\;
{\jv}(x)=(2mc)^{-1}\Psi^{\dagger a}_{\alpha }(x)
\stackrel{\longleftrightarrow}{{\Pop}}\Psi^a_{\alpha }(x),
\nonumber \\
&& \left\{\Psi^a_{\alpha }(x)\,,\Psi^b_{\beta}(y)\right\}\Biggr|_{x_0=y_0}=0,
\;\;\;\mbox{\bf with the convention:}
\nonumber \\
&& \left\{\Psi^a_{\alpha }(x)\,,\Psi^{\dagger b}_{\beta}(y)\right\}
\Biggr|_{x_0=y_0}=\delta_{\alpha\beta}\delta^{ab}\,\delta_3(\x-\y)
\Longrightarrow\Biggr|_{\x=\y}\delta_{\alpha\beta}\delta^{ab}\;\frac{1}{V^*}.
\nonumber
\eea
Here ${\cal E}(k)$ is arbitrary "bare" one-particle spectrum, $V^*$ has a
meaning of excitation volume, which could be connected with
momentum cut-off $\Lambda $: $6\pi^2/V^*=\Lambda^3$.
The second Hamiltonian is in\-va\-ri\-ant under (global)
$SU_I(2)\times SU_A(2)\times U(1)$ trans\-for\-ma\-ti\-ons
and corresponding generators can be found to be:
\bea
\left.\begin{array}{c}\hat{\rm I}^i \\ \hat{\rm A}^r\end{array}\;
\right\}=
\frac{1}{2}\int d^3{\rm x} \Psi^{\dagger a}_\alpha (x)\left\{
\begin{array}{c}(\sigma^i)_{\alpha\beta}\delta^{ab} \\
\delta_{\alpha\beta}(\sigma^r)^{ab} \end{array}\right\}\Psi^b_\beta (x);
\;\;\hat{\rm U} =
\frac{1}{2}\int d^3{\rm x} \Psi^{\dagger a}_\alpha (x)\Psi^a_\alpha (x),
\label{IRQ}
\eea
where $\hat{\rm I}^i$ are generators of isotopic $SU_I(2)$ transformations,
$\hat{\rm A}^r$ are  generators of additional $SU_A(2)$ transformations
and $\hat{\rm U}$ is $U(1)$ charge.
Introducing Heisenberg fields (HF) in momentum representation
$$
\Psi^a_{\alpha}(\x,t)=\int\frac{d^3k}{(2\pi)^{3/2}}\,e^{i(\ka\x)}\,
b^a_{\alpha}(\ka,t);\;\;
\left\{b^a_{\alpha}(\ka,t)\,,b^{\dagger b}_{\beta}(\q,t)\right\}=
\delta_{\alpha\beta}\delta_{ab}\delta_3(\ka-\q),
$$
we consider three different linear operator realizations of HF for $t=0$ via
physical fields, connected by Bogolubov rotations:
\bea
&& b^a_{\alpha}(\ka,0)=e^{\G}\,d^a_{\alpha}(\ka)\,e^{-\G}=
u_a\,d^a_{\alpha}(\ka)-v_a\epsilon_{\alpha\beta}d^{\dagger a}_{\beta}(-\ka),
\nonumber \\
&& \G=\frac{1}{2}\sum_{a=1,2}\vartheta^a\epsilon_{\alpha\beta}\int d^3k\left[
d^{\dagger a}_{\alpha}(\ka)d^{\dagger a}_{\beta}(-\ka)+
d^a_{\alpha}(\ka)d^a_{\beta}(-\ka)\right]=-\G^{\dagger},
\nonumber \\
&& u_a=\cos \vartheta^a;\;v_a=\sin \vartheta^a;\;(u_a)^2 + (v_a)^2=1,
\nonumber
\eea
which under condition $u_a v_a=0$ for $a=1,2$ lead to reduced Hamiltonians
in normal form exactly diagonalizable above relevant vacuums:
\bea
&& d^a_{\alpha}(\ka)\vac=0;\;\;
\left[H\{d\}\,,d^{\dagger a}_{\alpha}(\ka) \right]\vac =
E^a(k)\,d^{\dagger a}_{\alpha}(\ka)\vac ;\;\;H=Vw_0+\hat{H};
\label{dVac} \\
&& <{\cal E}(k)>\stackrel{def}{=}V^*\int \frac{d^3k}{(2\pi)^3}\,{\cal E}(k);
\;\; g=\frac{\lambda}{4V^*};\;\;\hat{H}\vac=0;
\;\;\hat{H}=\hat{H}_0+\hat{H}_I;
\nonumber \\
&& w_0=\frac{1}{V^*}\left[\left(2<{\cal E}(k)>-4g\right)
\left(v_1^2+v_2^2\right)-8g (v_1v_2)^2 \right];
\label{aW0} \\
&& E^a(k)=\frac{g}{(2mc)^2}\left(k^2+<k^2>\right)+g+
(1-2v_a^2)\left[{\cal E}(k)-2g(1+2v_{3-a}^2) \right].
\label{Eav} \\
&& \hat{H}_0\{d\}=\sum_{a=1,2}\di d^3k E^a(k)\,d^{\dagger a}_{\alpha}(\ka)
\,d^a_{\alpha}(\ka);
\label{aH0} \\
&& \hat{H}_I\{d\}=\frac{-1}{(2\pi)^3}\di d^3r \di d^3s \di d^3\PP
\Biggl\{\frac{1}{4}\left[\mu\left((\rv +\s)^2-\PP^2\right)+
\lambda\right]\cdot
\label{aHI} \\
&& \cdot\sum_{a=1,2}d^{\dagger a}_{\alpha}\left(\frac{\PP}{2}+\rv\right)
d^{\dagger a}_{\beta}\left(\frac{\PP}{2}-\rv\right)
d^a_{\beta}\left(\frac{\PP}{2}-\s\right)
d^a_{\alpha}\left(\frac{\PP}{2}+\s\right)+
\nonumber \\
&& +\frac{1}{2}\left[\mu\left( (\rv +\s)^2-\PP^2\right)+
\lambda (1-2v_1^2)(1-2v_2^2)\right]\cdot
\nonumber \\
&& \cdot d^{\dagger 1}_{\alpha}\left(\frac{\PP}{2}+\rv\right)
d^{\dagger 2}_{\beta}\left(\frac{\PP}{2}-\rv\right)
d^2_{\beta}\left(\frac{\PP}{2}-\s\right)
d^1_{\alpha}\left(\frac{\PP}{2}+\s\right) \Biggr\}; \quad
\mu=\frac{\lambda}{(2mc)^2}.
\nonumber
\eea
When $v_{1,2}$ take independently values 0,1 the diverse realizations
correspond to different systems: \\
1) B-system: $v_1=v_2=0$;
$d^1_{\alpha}(\ka)=B_{\alpha}(\ka)$; $d^2_{\alpha}(\ka)=\B_{\alpha}(\ka)$;
then $E^{1,2}_B(k)=E_B(k)$. Let $f^{ab}$ be an arbitrary
constant $SU_A(2)$ matrix, then the HF has the form:
\bea
&& \Psi^a_{\alpha}(x)\rightarrow \Psi^a_{\alpha}(x)_B=\di
\frac{d^3k}{(2\pi)^{3/2}}\left[f^{a1}{\cal B}_{\alpha}(\ka,t)+
f^{a2}\tilde{\cal B}_{\alpha}(\ka,t)\right]\,e^{i(\ka\x)-itE_B(k)};
\label{GFB} \\
&&  \left\{ \begin{array}{c}
{\cal B}_{\alpha}(\ka,t) \\ \tilde{\cal B}_{\alpha}(\ka,t)
\end{array} \right\} e^{-itE_B(k)}=e^{iHt}
\left\{\begin{array}{c} B_{\alpha}(\ka) \\ \B_{\alpha}(\ka) \end{array}
\right\} e^{-iHt};
\label{BGF}
\eea
One can check, that respective vacuum state $\vac_B$ is singlet for
both  $SU_I(2)$ and $SU_A(2)$ groups and the one-particle excitations
of $B$ and $\B$ form corresponding fundamental representations. \\
2) C-system: $v_1=v_2=1$;
$d^1_{\alpha}(\ka)=\epsilon_{\alpha\beta}C_{\beta}(\ka)$;
$d^2_{\alpha}(\ka)=\epsilon_{\alpha\beta}\C_{\beta}(\ka)$;
$E^{1,2}_C(k)=E_C(k)$,
\bea
\Psi^a_{\alpha}(x)\rightarrow \Psi^{\dagger a}_{\alpha}(x)_C=
\di \frac{d^3k}{(2\pi)^{3/2}}\left[f^{a1}{\cal C}^{\dagger}_{\alpha}(\ka,t)+
f^{a2}\tilde{\cal C}^{\dagger}_{\alpha}(\ka,t)
\right]e^{itE_C(k)-i(\ka\x)},
\label{GFC}
\eea
where the analogous to (\ref{BGF}) definitions are used.
The symmetry structure of this system is also similar to B-system. \\
3) A-system: $v_1=0$, $v_2=1$ (or vice versa).
It will be considered in detail. Let
$d^1_{\alpha}(\ka)=A_{\alpha}(\ka)$,
$d^2_{\alpha}(\ka)=\epsilon_{\alpha\beta}\A_{\beta}(\ka)$,
and let $f^{ab}$ be again an arbitrary constant $SU_A(2)$ matrix, then for
$ E^{2,1}_A(k)\equiv E^{(+,-)}_A(k)\equiv E_{\A,A}(k)$ respective HF
resemble relativistic ones:
\bea
\Psi^a_{\alpha}(x)_A=\int \frac{d^3k}{(2\pi)^{3/2}}
\left[f^{a1}{\cal A}_{\alpha}(\ka,t)e^{-itE^{(-)}_A(k)}+
f^{a2}\tilde{\cal A}^{\dagger}_{\alpha}(-\ka,t)e^{itE^{(+)}_A(-k)}\right]
e^{i(\ka\x)},
\nonumber \\
\mbox{where: }\left\{ \begin{array}{c}
{\cal A}_{\alpha}(\ka,t)\, e^{-itE^{(-)}_A(k)} \\
\tilde{\cal A}_{\alpha}(\ka,t)\, e^{-itE^{(+)}_A(k)}
\end{array} \right\} =e^{iHt}
\left\{\begin{array}{c} A_{\alpha}(\ka) \\ \A_{\alpha}(\ka) \end{array}
\right\} e^{-iHt}.\qquad \qquad\quad\qquad\qquad
\label{GFA}
\eea
Using (\ref{H2}), (\ref{IRQ}) and (\ref{GFA}) it is a simple
matter to show that for system A the $SU_A(2)$ and $U(1)$ symmetries turn
out to be spontaneously broken and there are four composite Goldstone states,
created by operators:
\bea
\hat{\G}^{\dagger}_{\alpha \beta}(0)\vac_A \equiv Z_0 \int d^3k\,
A^{\dagger}_\alpha(\ka)\,\A^{\dagger}_{\beta}(-\ka)\vac_A ,
\mbox{ satisfying to: } [H,\hat{\G}^{\dagger}_{\alpha \beta}(0)] = 0.
\label{Gld0}
\eea

\begin{center}
{\bf 3. Two-particle eigenvalue problems}
\end{center}

The interaction between all particles in the systems B and C is the same
as for $AA$, $\A\A$ in system A. So it is enough to consider the later
one. Hereafter $BB$ means $BB$, $\B\B$, $B\B$ and analogously for $CC$.
Defining two-particle interaction kernels and energies as:
\bea
&&
K^{\PP(QQ')}(\s,\ka)=\left\{\begin{array}{cc}
K^{\PP\{+\}}(\s,\ka), & \mbox{for} \quad QQ'= \A\A,AA,BB,CC \\
K^{\PP\{-\}}(\s,\ka), & \mbox{for} \quad QQ'= A\A \end{array}
\right.,
\nonumber \\
&&
-2K^{\PP\{\pm\}}(\s,\ka)=\frac{V^*}{(2\pi)^3}\cdot\frac{2g}{(2mc)^2}
\left[(\s+\ka)^2-\PP^2 \pm (2mc)^2\right],\;\mbox{ as well as:}
\nonumber \\
&&
E^{QQ'}_2(\PP,\ka)\equiv E_{Q}(\frac{\PP}{2}+\ka)+
E_{Q'}(\frac{\PP}{2}-\ka)= E^{\{\pm\} }_2(\PP,\ka),
\nonumber \\
&& E^{\{+\}}_2(\PP,\ka)=\frac{2g}{(2mc)^2}\left[<k^2> + (2mc)^2+k^2+
\frac{\PP^2}{4}\right]+
\label{E2+} \\
&& +\left\{\!\begin{array}{c}\displaystyle \!\!\!\!
\pm \left[4g-{\cal E}(\frac{\PP}{2}+\ka)-{\cal E}(\frac{\PP}{2}-\ka)\right]\\
\\ \displaystyle \,
\;\pm\left[12g-{\cal E}(\frac{\PP}{2}+\ka)-{\cal E}(\frac{\PP}{2}-\ka)\right]
\end{array}\! \right\}, \;\;\;\mbox{ for }
QQ'= \left\{\!\!\begin{array}{c}
\left[\!\!\begin{array}{c} \A\A \\  BB \!\!\end{array}\right] \\
\left[\!\!\begin{array}{c} CC \\  AA \!\!\end{array}\right]\\
\end{array}\!\!\right\};
\nonumber \\
&&
E^{\{-\}}_2(\PP,\ka)=\frac{2g}{(2mc)^2}\left[<k^2>- (2mc)^2+k^2+
\frac{\PP^2}{4}\right]+
\label{E2-} \\
&&
+\left[{\cal E}(\frac{\PP}{2}+\ka)-
{\cal E}(\frac{\PP}{2}-\ka)\right], \;\mbox{ for } \; QQ'=A\A;
\nonumber
\eea
we can formulate both scattering and bound state two-particle eigenvalue
problems in the Fock space created by the kinetic part of the reduced
Hamiltonian $\hat{H}$ (\ref{dVac}), (\ref{aH0}):
\be
\hat{H}\mid R^{\pm (QQ')}_{\alpha\beta}(\PP,\q)\rangle=E^{QQ'}_2(\PP,\q)\,
\mid R^{\pm (QQ')}_{\alpha\beta}(\PP,\q)\rangle ;\;\;
\hat{H}\mid {\rm B}^{\PP(QQ')}_{\alpha \beta} \rangle  =
M^{QQ'}_2(\PP)\mid {\rm B}^{\PP(QQ')}_{\alpha \beta} \rangle ;
\label{egnvl}
\ee
\[ \mid R^{\pm (QQ')}_{\alpha\beta}(\PP,\q)\rangle =\!\di d^3k\,
\Phi^{\pm QQ'}_{\PP q}(\ka)\mid R^{0(QQ')}_{\alpha\beta}(\PP,\ka)\rangle;
\,\;\mid {\rm B}^{\PP(QQ')}_{\alpha \beta}\rangle =\!\int d^3k\,
\Phi^{QQ'}_{\PP b}(\ka)\mid R^{0(QQ')}_{\alpha\beta}(\PP,\ka)\rangle;
\]
\be \mid R^{0(QQ')}_{\alpha\beta}(\PP,\ka)\rangle =
\hat{Q}^{\dagger}_{\alpha }(\frac{\PP}{2}+\ka)\,
\hat{Q}'^{\dagger}_{\beta }(\frac{\PP}{2}-\ka) \vac ;\qquad
M^{QQ'}_2(\PP)=E^{QQ'}_2(\PP,q=ib);
\label{egnst}
\ee
($\hat{Q},\hat{Q}'$ stands for creation operators $A,\A$, or $B,\B$, or
$C,\C$) in terms of Schr\"odinger equation for corresponding wave function:
\be
\left[E^{QQ'}_2(\PP,\ka)-M^{QQ'}_2(\PP)\right]\Phi^{QQ'}_{\PP}(\ka)=
-2\int d^3s\,\Phi^{QQ'}_{\PP}(\s)\, K^{\PP(QQ')}(\s,\ka).
\label{Prblm}
\ee
In order to search its finite renormalized solutions, let us
suppose at first only some general properties of "bare" parameters:
\bea
\lim_{k\rightarrow\infty}\left[{\cal E}(\frac{\PP}{2}+\ka)-
{\cal E}(\frac{\PP}{2}-\ka)\right]\cdot k^{-2}=0;\;\;{\cal E}(k)=mc^2 h(z^2);
\;\;z\equiv \frac{k}{mc};\;\; h'(0)<\infty;
\nonumber \\
\lim_{\Lambda\rightarrow\infty}\frac{2g}{(2mc)^2}=\frac{1}{{\cal M}_0}
\neq 0,\infty ;\qquad m=m(\Lambda)\sim\Lambda .\qquad\qquad
\nonumber
\eea
It is not hard to check that under these conditions eq.(\ref{Prblm}) has,
for the $\{-\}$ case, a simple solution independent from the very form of
"bare" spectrum:
\be
\Phi^{A\A}_{\PP}(\ka)=\Phi^{A\A}_{\PP}(0)\equiv Z_0;
\quad
M^{A\A}_2(\PP)=\frac{5}{4}\frac{2g}{(2mc)^2}\PP^2\Longrightarrow
\frac{5}{4{\cal M}_0}\PP^2,
\label{AGldst}
\ee
what performs obvious generalization on $\PP\neq 0$ of the Goldstone states
(\ref{Gld0})
\be
\hat{\G}^{\dagger}_{\alpha \beta}(\PP)\vac_A=Z_0\,\int d^3k\,
A^{\dagger}_\alpha(\frac{\PP}{2}+\ka)\,
\A^{\dagger}_{\beta}(\frac{\PP}{2}-\ka)\vac_A,
\label{GldP}
\ee
manifesting their Galileo invariance. The obtained solution (\ref{AGldst})
can be interpreted as bound state, in spite of the fact that its
configuration wave function is delta-function and cannot be normalized
by standard way. The relevant functional space will be specified bellow.
Nevertheless the corresponding pole is exactly canceled in T-matrix
(which vanishes with $\Lambda\rightarrow\infty $), what indicates that the
state (\ref{AGldst}) appears as a bound state embedded in continuum.
This collective Goldstone excitation (\ref{AGldst}), (\ref{GldP}) in
$A\A$ channel may be associated with spin-flip wave in some
ferromagnetic medium \cite{Umz}.

To discuss bound states of identical particles $\{+\}$ of $AA$ or $\A\A$ we
need more detailed form of "bare" spectrum. Supposing its quadratic form,
one can find from (\ref{dVac}), (\ref{Eav}) that in one particle sector:
\bea
{\cal E}(k)=\frac{k^2}{2m}+{\cal E}_0,\;\;\mbox{ is replaced by: }\;\;
E^{(\pm)}_A(k)=\frac{k^2}{2{\cal M}^{(\pm)}}+E^{(\pm)}_{A0};\;\mbox{ where,}
\label{bare} \\
\frac{1}{2{\cal M}^{(\pm)}}=\frac{g}{(2mc)^2}\mp \frac{1}{2m};\;\;\;
\lambda_0=\frac{\lambda {\cal M}^{(\pm)}}{2};\;\;\;
\mu_0=\frac{\lambda_0}{(2mc)^2}.
\label{mug}
\eea
For given $g(\Lambda)$, one can always choose the dependence
of "bare" parameters $m(\Lambda)$, ${\cal E}_0(\Lambda)$ on cut-off
$\Lambda$ leaving ${\cal M}(\Lambda)$, $E_0(\Lambda)$ finite when
$\Lambda\rightarrow\infty $, whereas $g(\Lambda)$ may be governed by
two-particle eigenvalue problem.
Really, when (\ref{bare}) is held the eq.(\ref{Prblm}) in configuration
space is reduced, for example for  "Toy" model (\ref{H1}), to the
Schr\"odinger equation with singular delta-potential
(here $c\rightarrow\infty$, ${\cal M}\rightarrow m $):
\be
\left(-\nabla^2_{\rm x}-q^2\right)\psi_q(\x)=\lambda_0\delta_3(\x)\psi_q(\x)
=\lambda_0\delta_3(\x)\psi_q(0).
\label{Dlt}
\ee
Its formal solutions \cite{BrFdd}
\bea
&& \VF{\ka}=\delta_3(\ka-\q)+\frac{\lambda_0\psi^{\pm}_q(0)}
{(k^2-q^2\mp i0)};
\qquad \Phi_{\PP b}(\ka)=\frac{\lambda_0 \psi_b(0)}{k^2+b^2};
\label{FB1}    \\
&& \lambda_0 \psi_b(0)=\sqrt{8\pi b};\;\;b>0;\;\;q=ib;
\nonumber \\
&& \left[\lambda_0\psi^{\pm}_q(0)\right]^{-1}= -2\pi^2(b\pm iq);\;\;\;
\mbox{sign}(b)\stackrel{instead}{\longrightarrow}\mbox{sign}(\lambda_0);
\label{bndcnd}
\eea
contain {\it arbitrary} real parameter $b$ of self-adjoint extension,
appearing in additional boundary conditions at the point $\x=0$
(\ref{bndcnd}), what are easily extracted from usual normalization
conditions of the formal solutions (\ref{FB1}):
\be
\int d^3k\, \Phi^{\pm *}_{{\cal P}p}(\ka)\,\Phi^{\pm}_{{\cal P}q}(\ka)=
\delta_3(\p-\q);\quad \int d^3k \,|\Phi_{\PP b}(\ka)|^2=1.
\label{nrmc1}
\ee
Indeed, e.c., for the scattering state one obtains
\be
\left[\lambda_0\psi^{\pm}_q(0)\right]^{-1}-
\left[\lambda^*_0\psi^{\pm *}_p(0)\right]^{-1}=-2\pi^2\left[\pm iq-
(\pm ip)^*\right],
\label{nrmc2}
\ee
that leads to (\ref{bndcnd}). The point is, since $\delta_3(\x)$ is
not an operator in ordinary sense , our quantum mechanical Hamiltonin
(\ref{Dlt}) is not well-defined on the whole Hilbert space but only on the
subspace $\psi(0)=0$ and needs some redefinition and extension. The
construction of well-defined self-adjoint extension for such singular
ill-defined operators needs {\it additional boundary conditions}
(\ref{bndcnd}) \cite{BrFdd},\cite{Hoeg-Horn}. At the same time, neither
$\lambda_0$ nor $\psi(0)$ have a meaning separately, but their product is
meaningful. On the other hand, the bound state equation (if it exists) with
cut-off prescription plays a role of transmutation condition of cut-off
$\Lambda$ and "bare" coupling constant into the unknown binding parameter $b$
\cite{Thorn}. In other words, the above self-adjoint extension under
using $\Lambda$-cut-off prescription implies the following subtraction
procedure with {\it fine-tuning} substitution \cite{BrFdd} for
\bea
&& \lambda_0\psi^{\pm}_q(0)=\frac{\lambda_0 (2\pi)^{-3}}{\Delta(\mp iq)};
\quad \Delta(\varrho)\equiv 1-\lambda_0\int \limits_{k<\Lambda}\frac{d^3k}
{(2\pi)^3}\frac{1}{(k^2+\varrho^2)}=1-\frac{\lambda_0}{2\pi^2}
\left(\Lambda-\frac{\pi}{2}\varrho\right):
\label{Delt} \\
&& \Delta(b)=0;\quad\Delta(\mp iq)\rightarrow\Delta(\mp iq)-\Delta(b);\quad
\lambda_0\rightarrow\tilde{\lambda}_0(\Lambda)\simeq\frac{2\pi^2}{\Lambda}+
\frac{\pi^3 b}{\Lambda^2}+\ldots .
\label{ftncnd}
\eea
From (\ref{bndcnd}), (\ref{ftncnd}) we notice that as well as in two-
dimensional case \cite{Thorn}, \cite{Sold}, there are no direct relationship
between the nature of point interaction and the sign of the quartic contact
self-interaction.
But the most important lesson is that we can formulate renormalization
conditions as additional boundary conditions (\ref{bndcnd}) without any
reference to $\Lambda$-cut-off regularization.

For the model (\ref{H2}) we face with the more singular potential,
studied in \cite{YShr}, \cite{Shond}, \cite{Fewst}:
\bea
\left(-\nabla^2_{\rm x}-q^2\right)\psi_q(\x)=\delta_3(\x)R_1(q)
-\nabla^2_{\rm x}\delta_3(\x)R_2(q)
-2\mu_0(\vec{\nabla}\psi_q)(0)\cdot\vec{\nabla}_{\rm x}\delta_3(\x);
\label{Scrd2} \\
R_1(q)\equiv (\lambda_0-\mu_0\PP^2)\psi_q(0)-\mu_0(\nabla^2\psi_q)(0);\;\;\;
R_2(q)\equiv\mu_0\psi_q(0).
\label{R12}
\eea
The last term in (\ref{Scrd2}) may be omitted for states with the orbital
momentum $l=0$ after integration over unit sphere of $\q$-directions.
As was shown in \cite{YShr}, construction of self-adjoint extension
for such strongly singular Hamiltonian generally requires the use of
space with redefined indefinite metric (belinear form) in order to include
wave functions with $r^{-1}$ and $\delta$ -singularities (already appeared
above for Goldstone state):
\bea
&& \phi(\x)=\varphi^s(r^2)+\frac{\varphi^a(r^2)}{r}+c_0\delta_3(\x)\equiv
\varphi^f(r)+c_0\delta_3(\x);
\nonumber \\
&& \widetilde{\int}d^3{\rm x}\phi^*(\x)\,\phi(\x)\stackrel{def}{=}
\int d^3{\rm x}\varphi^{f*}(r)\,\varphi^f(r)+
c_0\varphi^{s*}(0)+c^*_0\varphi^{s}(0);
\nonumber
\eea
($\varphi^{s,a}$-are regular functions near $r=0$),
and forthcoming narrowing it onto the invariant subspace with positively
defined metric. The narrowing conditions introduce again arbitrary real
parameters $\alpha_0,\beta$, marking corresponding self-adjoint extensions:
\be
c_0=4\pi\beta\varphi^{s}(0);\quad\varphi^a(0)=\alpha_0\varphi^s(0)-
\beta(\nabla^2\varphi^s)(0),
\label{Shcnd}
\ee
and give a simple solutions. For $l=0$ the solutions \cite{YShr} read
(for $\beta >0$ only one bound state exists):
$\tan\eta_q= q(\alpha_0+\beta q^2)$,
\bea
\phi^{(0)+}_q(\x)=
\frac{\sin(qr+\eta_q)}{qr}+4\pi\beta\cos\eta_q\delta_3(\x);\;\;\;
\phi^{(0)}_b(\x)=\frac{e^{-br}}{br}-4\pi\beta\delta_3(\x);
\nonumber \\
T^{\pm}(q)=\frac{1}{4\pi}\left[R^{\pm}_1(q)+q^2R^{\pm}_2(q)\right]=
\frac{e^{\pm i\eta_q}\sin\eta_q}{q}=\frac{\tan\eta_q}{q\,(1\mp i\tan\eta_q)};
\label{Shrsl} \\
\beta b^3-\alpha_0 b-1=0;\;\;\;
b=2\sqrt{\frac{\alpha_0}{3\beta}}\,\cos\frac{\tau}{3};\;\;\;
\cos\tau=\frac{1}{\sqrt{\sigma}};\;\;\;
\sigma=\frac{4}{\beta}\left(\frac{\alpha_0}{3}\right)^3,
\label{Shrbndst}
\eea
and satisfy for our case the following boundary-renormalization conditions,
generalizing (\ref{bndcnd}):
\bea
&& R^{\pm}_1(q)=\frac{4\pi\alpha_0}{(1\mp i\tan\eta_q)};\qquad
R^{\pm}_2(q)=\frac{4\pi\beta}{(1\mp i\tan\eta_q)};
\label{RR} \\
&& \frac{R_1(ib)}{R_2(ib)}\equiv (2mc)^2-\PP^2-
\frac{(\nabla^2\phi^{(0)}_b)(0)}{\phi^{(0)}_b(0)}=\frac{\alpha_0}{\beta}.
\nonumber
\eea
However, it seems impossible to obtain the last self-adjoint extensions in
the frameworks of $\Lambda$ cut-off regularization and so, impossible to
connect directly $\lambda_0,\mu_0,$ with $\alpha_0,\beta$.
Thus, $\alpha_0,\beta $ appear to be independent from $\PP$ that leads
to restoration of Galileo invariance for obtained solutions
(\ref{Shrsl}), (\ref{Shrbndst}). For $\beta=0$ we come back to results of
"Toy" model. For $l\geq 1$ the scattering and bound states are absent.

\begin{center}
{\bf 4. The $\Lambda$-cut-off prescription and fine tuning.}
\end{center}

In order to try to understand the second Hamiltinian on the language of
$\Lambda$ cut-off prescription let us put:
\bea
&& \gamma^{(\pm)}\equiv\frac{\mu\,{\cal M}^{(\pm)}}{2V^*}\equiv\frac{\mu_0}
{V^*}\equiv\frac{2{\cal M}^{(\pm)}g}{(2mc)^2}\equiv 1\pm
\frac{{\cal M}^{(\pm)}}{m};\;\;\;<k^2>=\frac{3}{5}\Lambda^2;
\label{gamma} \\
&&
g=\Lambda^2G(\Lambda);\;\;\;(2mc)^2=\Lambda^2\nu(\Lambda);\;\;\;
{\cal E}_0=\Lambda^2\epsilon(\Lambda);\;\mbox{ thus: }
\nonumber  \\
&& \gamma^{(\pm)}=\left[1\mp\frac{c\sqrt{\nu(\Lambda)}}{\Lambda
G(\Lambda)}\right]^{-1};\;
E^{(\pm)}_{A0}=\Lambda^2\left\{G(\Lambda)\left[
\frac{3}{5\nu(\Lambda)}-1\pm 4\right]\mp\epsilon(\Lambda)\right\}.
\label{Cutof} \\
&& \mbox{Further: }
G(\Lambda)=G_0+\frac{G_1}{\Lambda}+\frac{G_2}{\Lambda^2}+\ldots;
\mbox{ and so for: }\nu(\Lambda),\epsilon(\Lambda),\gamma(\Lambda).
\label{nuGLda}
\eea
For $G_0,\nu_0\neq 0$ the following quantities should be finite:
\bea
&&\gamma^{(\pm)}_0=\gamma_0=1;\;\;\;\gamma^{(\pm)}_1=
\pm c\frac{\sqrt{\nu_0}}{G_0};\;\;\;
{\cal M}^{(\pm)}_0={\cal M}_0=\frac{\nu_0}{2G_0};
\label{01}  \\
&& E^{(\pm)}_{A0}=\Lambda^2\left[G_0\left(\frac{3}{5\nu_0}-1\pm 4\right)\mp
\epsilon_0\right]+\Lambda\left[G_1\left(\frac{3}{5\nu_0}-1\pm 4\right)
\mp\epsilon_1 -\frac{3G_0\nu_1}{5\nu^2_0}\right]+
\nonumber  \\
&& +\left[G_2\left(\frac{3}{5\nu_0}-1\pm 4\right)\mp\epsilon_2-\frac{3}{5}
\left(\frac{G_0\nu_2+G_1\nu_1}{\nu^2_0}-\frac{G_0\nu^2_1}{\nu^3_0}\right)
\right].
\label{012}
\eea
Thus, the renormalization conditions in one-particle sector imply that
the expressions in first and second square brackets (\ref{012}) vanish.
The demand of finiteness of {\it both} one-particle spectra
$E^{(\pm)}_A(k)$ of A-system leads to relations (for B,C-systems only the
conditions following from analog of (\ref{012}) take place):
\bea
&& \nu^A_0=\frac{3}{5};\quad {\cal M}_0=\frac{0.3}{G_0};\quad
\nu^A_1=0;\quad \epsilon^A_{0,1}=4G_{0,1};\quad \mbox{then :}
\label{Acnd} \\
&& E^{(\pm)}_A(k)= \frac{k^2-\nu_2}{2{\cal M}_0}\pm
(4G_2-\epsilon_2),\quad (\mbox{so here } \nu_0,G_0>0).
\label{AE1rn}
\eea
To write the formal solution of (\ref{Prblm}) in momentum representation
for {\it arbitrary} "bare" spectra we define:
\bea
&& \left\{\begin{array}{c}
\J{n} \\ \J{D} \\ \J{jl} \end{array}\right\}=
\frac{1}{2} \di \frac{d^3r}{(2\pi)^3}\,\cdot \frac{1}{\E{r}-i\sigma}\cdot
\left\{\begin{array}{c} \displaystyle\left(r^2\right)^n   \\
\displaystyle \frac{\left(\rv\cdot\PP\right)^2}{\PP^2}  \\
\displaystyle  r^j r^l \end{array}\right.  ;
\label{JJJ} \\
&& \J{jl}=\delta_{jl}\J{\delta}+\frac{\PP^j\PP^l}{\PP^2}\J{R},\quad
\J{\delta}=\frac{1}{2}\left[\J{1}-\J{D}\right];
\nonumber \\
&& \left.\begin{array}{c}\D{0} \\  \D{\kappa \,1} \\
 \D{\kappa \,2}  \end{array}\right\}= \left\{
\begin{array}{c}  \displaystyle
\left[\mu\,\J{1}-1\right]^2- \mu^2\,\J{0}\J{2}-(\lambda-\mu\PP^2)\,\J{0}.\\
\displaystyle
\lambda-\mu\PP^2+\mu \kappa^2+\mu^2\left[\J{2}-\kappa^2\J{1}\right]. \\
\displaystyle
\mu +\mu^2 \left[\kappa^2\J{0}-\J{1}\right].
\end{array} \right. .
\label{DDD}
\eea
Setting $\I{\{\ldots\}}= {\rm J}^{\cal P}_{\{\ldots\}}(\pm\delta-i\E{q}),
\quad \DD{\{\ldots\}}=D^{\cal P}_{\{\ldots\}}(\pm\delta-i\E{q})$,
for the scattering eigenfunctions with fixed parity (angular momentum)
$l=0,1$ and spin J=0,1, defined in symmetrical basis
($\sigma_j,\; j=1,2,3$ -usual Pauli matrices):
\bea
(\delta_{\alpha \beta},(\sigma_j)_{\alpha \beta})\longrightarrow
(\epsilon_{\alpha \beta},\tau^j_{\alpha \beta}), \;\;\;
\epsilon_{\alpha \beta}=i(\sigma_2)_{\alpha \beta},\;\;\;
\tau^j_{\alpha \beta}=\tau^j_{\beta\alpha}=
i(\sigma_2\sigma_j)_{\alpha \beta}, \quad\mbox{ as:}
\label{tau} \\
\mid l,{\rm J,m};\PP,q\rangle^{\pm}=\di d^3k\,\vf{l,\rm J,m}{\ka}\,
\mid R^{0(QQ')}_{\alpha\beta}(\PP,\ka )\rangle;\;\;\;
\vf{l,\rm J,m}{\ka}=\chi^{(\rm J,m)}_{\alpha \beta}\,
\Phi^{\pm(l)}_{{\cal P}q}(\ka);
\label{2states} \\
\mbox{where for } Q=Q':\quad
\vf{l,\rm J,m}{\ka}=-\phi^{\pm (l,{\rm J,m})}_{{\cal P}q}
(-\ka)_{\beta\alpha};\; \mbox{ and then }\;l={\rm J}=0,1;
\nonumber \\
\tau^{(0)}_{\alpha \beta}=i\tau^{3}_{\alpha \beta};\;\;\;
\tau^{(\pm 1)}_{\alpha \beta}=\frac{\mp i}{\sqrt{2}}\left(
\tau^{1}_{\alpha \beta}\pm i\tau^{2}_{\alpha \beta}\right);
\qquad \chi^{(0,0)}_{\alpha \beta}=
\epsilon_{\alpha \beta};\;\;\;\chi^{(1,\rm m)}_{\alpha \beta}=
\tau^{(\rm m)}_{\alpha \beta};
\nonumber
\eea
one has the following expressions:
\bea
&& \Phi^{\pm(l)}_{{\cal P}q}(\kap)=\frac{1}{2}\left[\delta_3(\kap-\q)+
(-1)^{l} \delta_3(\kap+\q)\right]+
\frac{\TT{l}{\kappa} }{\E{\kappa}-\E{q}\mp i0 };
\nonumber \\
&& \TT{0}{k}=\CC{1}+k^2\,\CC{2}=\frac{ \DD{q1}+k^2\,\DD{q2} }
{2(2\pi)^3 \DD{0} },
\label{TT0} \\
&& \TT{1}{k}= \left(\ka \cdot \CC{3}\right) =
\frac{\mu }{(2\pi)^3} k^j\left\{
\frac{{\prod}^{jl}_{\perp}(\PP)}{ 1-2\mu \I{\delta} }+
\frac{{\prod}^{jl}_{\parallel}(\PP)}{ 1-2\mu \I{D} } \right\} q^l;
\nonumber \\
&& \mbox{where projectors are}\;\;
{\prod}^{jl}_{\perp}(\PP)=\left(\delta_{jl}-\frac{\PP^j\PP^l}{\PP^2}
\right); \quad {\prod}^{jl}_{\parallel}(\PP)=\frac{\PP^j\PP^l}{\PP^2}.
\label{TT1}
\eea
The bound state wave functions look like simple residues of scattering ones
$\Phi^{\pm(l)}_{{\cal P}q}(\ka)$ for corresponding poles for
$\E{q}\pm i0\rightarrow i\sigma \rightarrow M^{(l)}_2(\PP)$,
$q\rightarrow ib(\PP)$,
\bea
&& \D{0}=0; \quad 1-2\mu \J{\delta}=0; \quad 1-2\mu \J{D}=0;
\label{bndst} \\
&&\Phi^{(l)}_{{\cal P}b}(\ka))=Z^{(l)}(\PP,\ka)\left[\E{k}-M^{(l)}_2(\PP)
\right]^{-1};\;\;{\rm C}_n(\PP)=\mbox{ res }\CC{n};
\nonumber \\
&& Z^{(0)}(\PP,\ka)={\rm C}_1(\PP)+k^2\,{\rm C}_2(\PP);\quad
Z^{(1)}_{\perp,\parallel}(\PP,\ka)=\left(\ka \cdot
\vec{\rm C}^{\perp,\parallel}_3(\PP)\right);
\nonumber
\eea
Let the bound state equation (\ref{bndst}), for orbital momentum $l=0$,
spin J=0, as well as for the "Toy" model (\ref{ftncnd}), serve as a
dimensional transmutation condition. For particles with identical quadratic
spectra (\ref{bare}), with the help of the simple relations for integrals
(\ref{JJJ}):
\bea
&&\mu{\rm J}^{\PP}_n\left(-iE_2(\PP,i\varrho)\right)\Longrightarrow
\JJ_n(\varrho)\equiv\gamma V^*\,\di\limits^{\Lambda}\frac{d^3k}{(2\pi)^3}
\cdot\frac{\left(k^2\right)^n}{(k^2+\varrho^2)};
\label{J210} \\
&& \JJ_0(\varrho)=\frac{\gamma V^*}{2\pi^2}
\left(\Lambda-\varrho\arctan\frac{\Lambda}{\varrho}\right)
=\frac{\gamma V^*}{2\pi^2}\left(\Lambda-\frac{\pi}{2}\sqrt{\varrho^2}+
\varrho\arctan\frac{\varrho}{\Lambda}\right);
\nonumber \\
&& \JJ_1(\varrho)=\gamma -\varrho^2\,\JJ_0(\varrho);\;\;\;\;
\JJ_2(\varrho)=\gamma <k^2> -\varrho^2\,\JJ_1(\varrho);
\nonumber \\
&& M_2(\PP)\equiv \E{q=ib}=\frac{\PP^2}{4{\cal M}^{(\pm)}}+2E^{(\pm)}_0-
\frac{b^2}{{\cal M}^{(\pm)}},
\label{M2E}
\eea
it reads:
\bea
&& {\cal D}^{\PP}_0(\varrho)\equiv (\gamma-1)^2-\JJ_0(\varrho)\left[(2mc)^2+
\gamma <k^2>-\PP^2-(2-\gamma)\varrho^2\right]\Longrightarrow
\nonumber  \\
&& \Longrightarrow (\gamma-1)^2-\gamma(2-\gamma)\left(Y-z^2 \right)
\left(1-z\arctan\frac{1}{z}\right)\rightarrow 0;\;\;\mbox{ for: }
\varrho\rightarrow b;
\label{DMb} \\
&& b=\Lambda z;\;\;\;Y=\frac{(3/5)\gamma+\tilde{\nu}}{2-\gamma};\;\;\;
\tilde{\nu}(\Lambda)=\nu (\Lambda)-\frac{\PP^2}{\Lambda^2};\;\;\;
Y(\Lambda)=Y_0+\frac{Y_1}{\Lambda}+\ldots .
\label{XYGb}
\eea
Let us point out, that the limit $\Lambda\rightarrow\infty$ must be
performed here {\it after} carrying out $k$-integrations. Now the
finiteness of $b$ implies the existence of solutions $z_0=0$ what,
together with (\ref{01}), (\ref{012}), gives the following fine-tuning
relations:
\bea
&& Y_0=0;\;\;\;\nu_0=-\frac{3}{5}\gamma_0=-\frac{3}{5};\;\;\;
{\cal M}_0\equiv\frac{\nu_0}{2G_0}= -\frac{0.3}{G_0};\;\;\;
(\nu_0,\,G_0<0);
\nonumber \\
&& \gamma^{(\pm)}_0=1;\;\;\;\gamma^{(\pm)}_1=\pm i\frac{c}{G_0}
\sqrt{\frac{3}{5}};\;\;\;Y_1=\frac{3}{5}\gamma_1+\nu_1\equiv
\frac{2}{\pi}\Upsilon;
\label{rnrmz} \\
&& E_{\left\{\!\!\!\tiny\begin{array}{c} \A \\ B \\ C \\ A \end{array}\!\!\!
\right\}}(k)=\frac{k^2+\nu_2}{2{\cal M}_0}-\frac{5}{3}\nu_1
\left(G_1-\frac{\nu_1}{2{\cal M}_0}\right)+\left\{\begin{array}{c}
\displaystyle \pm [2G_2-\epsilon_2]\\ \\ \displaystyle
\pm [6G_2-\epsilon_2] \end{array}\right\};
\label{E1rn} \\
&&\mbox{for: }\;\;\;\;\;\left\{\begin{array}{c}\displaystyle \left[
\epsilon_0=2G_0;\;\;\;\epsilon_1=2G_1\pm\frac{\nu_1}{2{\cal M}_0}\right] \\
\\ \displaystyle
\left[\epsilon_0=6G_0;\;\;\;\epsilon_1=6G_1\pm\frac{\nu_1}{2{\cal M}_0}
\right]\end{array}\right\},\quad \mbox{ correspondingly.}
\label{fntun1}
\eea
The renormalized half-off-shell T-matrix for $l=0$ (\ref{TT0}) after simple
subtraction in denominator: $ \DD{0}\rightarrow\DD{0}-{\cal D}^{\PP}_0(b)$,
looks:
\[-\frac{1}{2\pi^2{\cal M}}\cdot\frac{
\Lambda^2Y+\left[q^2+k^2(1-\gamma)\right]/(2-\gamma)}
{\Lambda^2Y(b\pm iq)+(2/\pi)\Lambda (1-Y)(b^2+q^2)-
\left(b^3\pm (iq)^3\right) +O(1/\Lambda)},
\]
and under the fine-tuning conditions (\ref{rnrmz}) takes the form :
\be
\TT{0}{k}\Bigr|_{\Lambda\rightarrow\infty}=-\frac{1}{2\pi^2{\cal M}_0}
\cdot \frac{\Upsilon}{(\Upsilon+b\mp iq)(b\pm iq)}.
\label{T0rnrm}
\ee
Moreover, it is not hard to check the same subtraction procedure leads the
same limiting expression for {\it off-shell} T-matrix:
\bea
&& 2\pi^2{\cal M}\,\langle\s\ |\ \hat{\rm T}(-\varrho^2)\ |\ \rv\rangle =
\frac{\gamma V^*}{4\pi}\left[\frac{{\cal O}^{\PP}(\varrho;\s,\rv)}
{{\cal D}^{\PP}(\varrho)}+\frac{2(\s\cdot\rv)}{1-\frac{2}{3}\gamma+
\frac{2}{3}\varrho^2\ \JJ_0(\varrho)}\right];
\label{offsT} \\
&&{\cal O}^{\PP}(\varrho;\s,\rv)\equiv\gamma <k^2>+(2mc)^2-\PP^2-\varrho^2+
\nonumber \\
&& +(1-\gamma)(s^2+r^2+\varrho^2)+\JJ_0(\varrho)(s^2+\varrho^2)
(r^2+\varrho^2).
\nonumber
\eea
Corresponding bound state wave function with $l=0$ has the same form as for
the "Toy" case (\ref{FB1}). Note that for the case at hand the Galileo
invariance which means independence on $\PP$ of both scattering and bound
state eigenfunctions again is restored manifestly only due to the applied
renormalization procedure. The same result (\ref{T0rnrm}) takes place also
for B and C systems. Self consistent boundary-renormalization conditions,
generalizing the (\ref{bndcnd}), for this solution may be chosen as
following:
\be
\lambda_0\psi_q(0)=\mu_0\psi_q(0)=0;\;\;
\mu_0(\nabla^2\psi^{\pm}_q)(0)=
\frac{1}{2\pi^2}\cdot\frac{\Upsilon}{(\Upsilon+b\mp iq)(b\pm iq)}.
\label{bndr}
\ee
For $l=1$ T-matrix (\ref{TT1}) becomes:
\be
\TT{1}{k}=\frac{\mu}{(2\pi)^3}\cdot\frac{(\ka\cdot\q)}{1-\frac{2}{3}\mu
\I{1}},
\label{T1}
\ee
and for $\gamma_0=1$ it tends to zero like $\Lambda^{-3}$ when
$\Lambda\rightarrow\infty$. So, there are no any scattering and bound
states with $l=1$ for such extension, determined
by (\ref{Cutof}), (\ref{nuGLda}), (\ref{01}), (\ref{rnrmz}),
(\ref{fntun1}).

To find another extension let us apply the above described
renormalization procedure directly to the case $l=1$. The denominator of
(\ref{T1}) with $q\rightarrow ib_1$, $b_1=\Lambda y$ gives the
transmutation conditions of the type:
$y^3-3\xi y^2-\xi\left(3/(2\gamma)-1\right)=0$, $y_0=0$,
which mean that $\gamma_0=3/2$. Then the finiteness of ${\cal M}_0$
leads to the requirements: $G_0=G_1=\nu_0=\nu_1=0$.
However, for this case both T-matrices $\TT{0}{k},\;\TT{1}{k}$ tend to
zero like $\Lambda^{-1}$.
So, such extension is completely equivalent to the free Hamiltonian.

The second possibility is to choose the finite "bare" mass. Thus
$G_{0,1}=\nu_{0,1}=0$,
\bea
&& g(\Lambda)=g_0+\frac{g_1}{\Lambda}+\ldots ;\;\;
g_n=G_{n+2};\;\;\;{\cal M}^{(\pm)}=\pm m(\gamma^{(\pm)}-1);
\label{gLmd} \\
&& E^{(\pm)}_{A0}(\Lambda)=
\Lambda^2\left[\frac{3g_0}{5(2mc)^2}\mp \epsilon_0\right]+
\Lambda\left[\frac{3g_1}{5(2mc)^2}\mp \epsilon_1\right]+
\nonumber     \\
&& +\left[\frac{3g_2}{5(2mc)^2}-g_0\pm \left(4g_0-\epsilon_2\right)\right].
\label{E0m}
\eea
Here the first and second square brackets must vanish again, but
this could not be fulfilled for both spectra in A-system simultaneously.
By the same way as above we come to the transmutation condition for the
case $(-)$ of $A$ -spectrum :
\bea
&& {\cal D}^{\PP}_0(b)\mid_{\Lambda\rightarrow\infty}=
(\gamma_0-1)^2-\frac{9}{5}\gamma^2_0=0;\quad\mbox{ then: }
\label{D0m} \\
&& \gamma_0=1-\frac{3}{4}(3\pm\sqrt{5})\equiv\gamma^{(-)}_0 <1;\;\;\;
{\cal M}^{(-)}_0=m\frac{3}{4}(3\pm\sqrt{5});\;\;\;
b=\frac{4}{\pi}\frac{g_1}{g_0}.
\label{rnmmcnd}
\eea
Now the scattering and bound state solutions look the same as for the "Toy"
model (\ref{FB1}):
\be
\TT{0}{k}\mid_{\Lambda\rightarrow\infty}=
-\frac{1}{2\pi^2{\cal M}^{(-)}_0}\cdot\frac{1}{(b\pm iq)};\;\;\;
\phi^{(0)}_b(\x)=\frac{\sqrt{8\pi b}}{4\pi}\frac{e^{-br}}{r};\;\;r=|\x|.
\label{TB0m}
\ee
For $\A\;(+)$ case such kind of renormalization and bound state are
impossible if $m$ remains finite. The scattering and bound state for
$l=1$ here is absent again.
The same results are also true for B and C systems. Besides B-system
corresponds here to the $(-)$ case, whereas C-system, to the $(+)$ one.
Obviously, we recognize here the extension from the previous section with
$\beta =0$, however the respective general case (\ref{Shrsl}),
(\ref{Shrbndst}), (\ref{RR}) is reproduced by the formulas (\ref{T0rnrm}) -
(\ref{T1}) only qualitatively: they give only the same number of bound states
and absence of any interaction for $l=1$, but expressions (\ref{T0rnrm}) and
(\ref{Shrsl}) for T-matrix with $l=0$ are essentially different.
\newpage

\begin{center}
{\bf 5. Discussion}
\end{center}

The conditions (\ref{rnrmz}), (\ref{fntun1}), (\ref{E1rn}) are opposite
to (\ref{Acnd}), (\ref{AE1rn}). So, in the frameworks of $\Lambda$-cut-off
prescription in A-system it seems impossible to have renormalized bound
(and scattering) eigenstates of the identical particles and finite spectra
for particles of both type simultaneously. However, for B and C-systems the
both one- and two-particle sectors are simultaneously well defined by this
way. On the other hand, there are no obstacles for A-system also to derive
them in the form (\ref{T0rnrm}), (\ref{TB0m}), as well as (\ref{Shrsl}),
(\ref{Shrbndst}), by the methods of theory of self-adjoint extension in the
appropriate functional space \cite{YShr}, \cite{Shond}, \cite{Fewst}.
Really, it was point out by Berezin \cite{Ber}, that (for three-dimensional
case) the $\Lambda$- cut-off prescription may be used only to pick out the
unique extension among the different possible ones.
As was shown in \cite{Shond} the solutions (\ref{T0rnrm}), (\ref{TB0m}),
(\ref{Shrsl}) realize {\it different} self-adjoint extensions corresponding
to operator (\ref{Scrd2}), but the $\Lambda$- cut-off prescription now
separates only (\ref{T0rnrm}) or (\ref{TB0m}).
In any case the real $\alpha_0$ and $\beta$ or $\Upsilon$ and $b$ are
{\it arbitrary} parameters of self-adjoint extension which may be formally
partially expressed via parameters of formal $\Lambda$-dependence of "bare"
quantities by fine tuning relations (\ref{01}), (\ref{rnmmcnd}).
It is easy to show similarly to (\ref{nrmc1}), (\ref{nrmc2}), that,
strictly speaking, the solutions (\ref{Shrsl}), (\ref{T0rnrm}),
imply a self-adjoint extension for restricted on relevant subspace of $L^2$
initial free Hamiltonian to {\it extended Hilbert space} $L^2\oplus C_1$
with different structure of additional space $C_1$.
The additional discrete component of eigenfunctions "improves" their scalar
product, it is completely defined by the same parameters of self-adjoint
extension but does not affect on physical meaning of obtained solution in
ordinary space \cite{Shond}, \cite{Fewst}.

\begin{center}
{\bf 6. Conclusions}
\end{center}

So, we have formulated several unambiguous renormalization procedures to
extract a renormalized dynamics from "nonrenormalizable" contact four-fermion
current-current interaction that are selfconsistent in one- and two-particle
sector and intimately connected with construction of self-adjoint
extensions of corresponding quantum mechanical Hamiltonian and restoration
of Galileo invariance. The point is that additional boundary conditions
determining the self-adjoint extension in two-particle sector play a role
of renormalization conditions. The $\Lambda$-cut-off prescription having
one renormalization for that problems is explicitly demonstrated.

We find three unitary - inequivalent operator realizations of four-
fermion Hamiltonian in which the two-particle eigenstates' problem can
be studied and solved exactly, and show that spontaneous symmetry breaking
and existence of corresponding Goldstone modes are inalienable properties
of fermion current-current interaction for vector current with rich enough
symmetry.

The problems discussed here for nonrelativistic field theories could take
place also for relativistic ones. So, it seems reasonable
that the renormaliziation procedure for relativistic case also
may be meaningful beyond perturbation theory and may be formulated
like additional boundary condition for solutions of Bete-Salpeter
equation.
However, it is well-known \cite{Savr}, that formulation of such conditions
is a separate unsolved problem of quantum field theory.

The authors are grateful to A.A.Andrianov, R.Soldati and L.D.Faddeev for
useful discussions and V.L.Chernyak and I.B.Khriplovich for valuable remarks.

\end{document}